\documentclass[conference]{IEEEtran}
\IEEEoverridecommandlockouts

\usepackage{cite}
\usepackage{amsmath,amssymb,amsfonts}
\usepackage{algorithm}
\usepackage{algpseudocode}
\usepackage{graphicx}
\usepackage{textcomp}
\usepackage{mathtools}
\usepackage{multirow}
\usepackage{bm}
\usepackage[dvipsnames]{xcolor}
\def\BibTeX{{\rm B\kern-.05em{\sc i\kern-.025em b}\kern-.08em
    T\kern-.1667em\lower.7ex\hbox{E}\kern-.125emX}}
    
\begin{document}

\title{Joint Optimisation of Electric Vehicle Routing and Scheduling: A Deep Learning-Driven Approach for Dynamic Fleet Sizes\\
{\footnotesize}
\thanks{Hao Wang is supported in part by the Australian Research Council Discovery Early Career Researcher Award under Grant DE230100046.}}

\author{
    \IEEEauthorblockN{Jun Kang Yap\IEEEauthorrefmark{1}, Vishnu Monn Baskaran\IEEEauthorrefmark{1}, Wen Shan Tan\IEEEauthorrefmark{2}, Ze Yang Ding\IEEEauthorrefmark{3}, Hao Wang\IEEEauthorrefmark{4}, David L. Dowe\IEEEauthorrefmark{4}}
    \IEEEauthorblockA{\IEEEauthorrefmark{1}School of Information Technology \IEEEauthorrefmark{2}\IEEEauthorrefmark{3}School of Engineering \IEEEauthorrefmark{2}Centre for Net-Zero Technology
    \\ \IEEEauthorrefmark{1}\IEEEauthorrefmark{2}\IEEEauthorrefmark{3}Monash University Malaysia, Bandar Sunway, Malaysia
    \\ \IEEEauthorrefmark{1}\IEEEauthorrefmark{2}\IEEEauthorrefmark{3}\{jun.yap, vishnu.monn, tan.wenshan, ding.zeyang\}@monash.edu}
    \IEEEauthorblockA{\IEEEauthorrefmark{4}Department of Data Science and AI, Faculty of IT
    \\ \IEEEauthorrefmark{4}Monash University, Clayton, Australia
    \\ \IEEEauthorrefmark{4}\{hao.wang2, david.dowe\}@monash.edu}
}

\maketitle

\begin{abstract}
Electric Vehicles (EVs) are becoming increasingly prevalent nowadays, with studies highlighting their potential as mobile energy storage systems to provide grid support. Realising this potential requires effective charging coordination, which are often formulated as mixed-integer programming (MIP) problems. However, MIP problems are NP-hard and often intractable when applied to time-sensitive tasks. To address this limitation, we propose a deep learning assisted approach for optimising a day-ahead EV joint routing and scheduling problem with varying number of EVs. This problem simultaneously optimises EV routing, charging, discharging and generator scheduling within a distribution network with renewable energy sources. A convolutional neural network is trained to predict the binary variables, thereby reducing the solution search space and enabling solvers to determine the remaining variables more efficiently. Additionally, a padding mechanism is included to handle the changes in input and output sizes caused by varying number of EVs, thus eliminating the need for re-training. In a case study on the IEEE 33-bus system and Nguyen-Dupuis transportation network, our approach reduced runtime by 97.8\% when compared to an unassisted MIP solver, while retaining 99.5\% feasibility and deviating less than 0.01\% from the optimal solution.
\end{abstract}


\section{Introduction} \label{introduction}
The worsening impacts of global warming have intensified efforts to reduce carbon emissions. Since fossil fuel vehicles are one of the major contributors, this has prompted a growing shift towards the adoption of electric vehicles (EVs). This trend can been observed from the 2 million EV registrations in Europe in 2022 alone \cite{10751406}. In addition to reducing carbon emissions, studies have shown that EVs can act as mobile energy storage systems (ESSs) to help improve electric power grid stability and renewable energy source (RES) utilisation \cite{10.1145/3678717.3691247}. This is because EVs acting as mobile ESSs can potentially help redistribute surplus RES generation to regions with higher demand at a lower cost via the transportation network (TN). However, realising these benefits hinges on the development of an efficient yet effective charging coordination strategy. 

\begin{figure}[!t]
\centerline{\includegraphics[width=\linewidth]{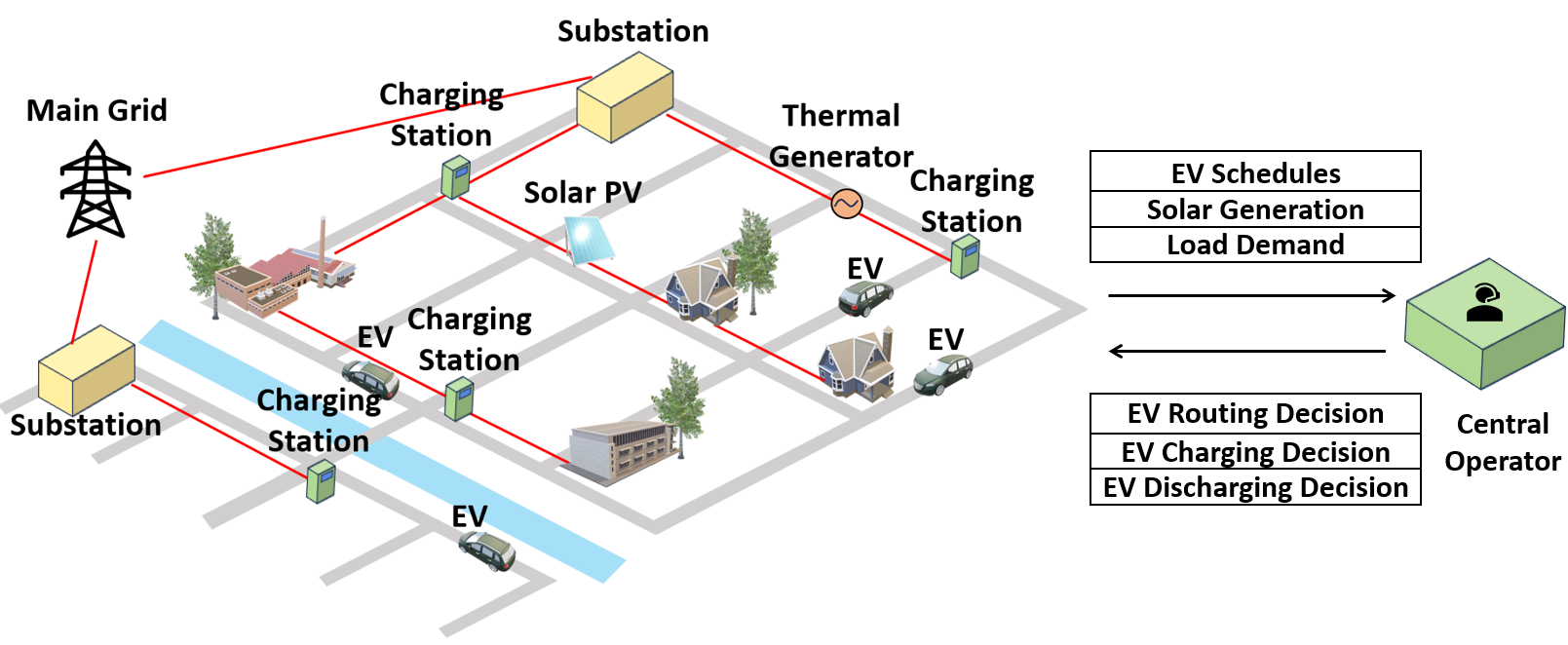}}
\caption{A variation of an EV charging coordination problem.}
\label{coordination}
\end{figure}

There are a wide range of formulations for the EV charging coordination optimisation problem and Fig. \ref{coordination} shows an example of it. These formulations can generally be classified into two main categories: those that focus exclusively on grid side optimisation and those that incorporate TNs. The former has been demonstrated in \cite{rafique2021ev}, where the authors formulated an EV charging coordination model to minimise the energy purchased from the main grid for a distribution network (DN) with solar photovoltaic (PV) systems. Similarly, \cite{rafique2022two} proposed a model to minimise the energy purchased to charge the electric buses in a depot equipped with solar PVs. Moreover, \cite{9540895} introduced a tri-level strategy to maximise EV aggregators' charging profits. Additionally, \cite{yang2019binary} and \cite{yang2015unit} explored a large EV fleet charging coordination problem with the unit commitment (UC) problem. However, focusing solely on grid side optimisation may not be appropriate as charging stations introduce a coupling relationship between the grid and TN, and poor coordination on either side can adversely impact the other.

Recent studies have begun exploring this coupling relationship by incorporating TN in their formulations. Instead of forecasting or modelling EV arrival-departure times, and state of charge at charging stations as uncertainties, earlier research solves a vehicle routing problem to determine these parameters. The returned routing decisions are then used to perform grid-side scheduling. This approach has been demonstrated in \cite{aghajan2023optimal, liu2022collaborative, yao2021joint}. However, according to \cite{TRIVINOCABRERA2019113}, better overall optimisation performance may be achieved if routing and scheduling were to be jointly optimised, as grid conditions would be made visible during routing. Therefore, a mixed-integer programming (MIP) model was formulated in \cite{TRIVINOCABRERA2019113} to jointly optimise EV routing and charging-discharging to maximise EV charging profits, while \cite{zhao2022congestion} extended this formulation by including queuing times at charging stations. Despite that, these formulations often still face challenges in representing varying travel times throughout the scheduling period due to discrepancies in time-resolution between the routing and scheduling side. Therefore, our work aims to address this with the addition of virtual congestion nodes (VCNs), which enables dynamic adjustment of trip times in the TN throughout the scheduling period.

The joint routing and scheduling (JRS) EV charging coordination problem can often be formulated as a MIP problem. However, the combinatorial complexity introduced by the integer variables can render them impractical for applications with strict time constraints, such as the 3–4 hour clearing window in the day-ahead energy trading market \cite{7419921}. Thus, this has driven researchers to explore  alternatives to expedite MIP's solution process. Recently, machine learning (ML), particularly deep learning (DL), has shown promising potential in solving power-system optimisation problems due to its fast inference time and capability to learn complex functions. For example, \cite{9810496} applied feed-forward networks (FFNs),  convolutional neural networks (CNNs) and graph neural networks to solve an optimal power flow (OPF) problem by predicting the generator set points. Similarly, \cite{dinh2022supervised} and \cite{kim2020supervised} had also used FFNs to solve MIP energy management problems. Although substituting MIP solvers with ML can result in significant acceleration, ensuring solution feasibility is often challenging as ML models cannot enforce problem constraints. 

To address this, researchers have instead explored hybrid approaches to leverage solver’s ability to handle constraints. These approaches typically involve simplifying MIP problems with ML, thereby allowing them to be solved more efficiently by solvers. For instance, a FFN had been trained in \cite{jovanovic2020online} to predict a partial solution to assist a heuristic algorithm in solving a charging station assignment problem. Moreover, \cite{9810496} and \cite{9454298} reduced the number of constraints in OPF problems by using ML to predict the active set constraints. Besides that, k-nearest neighbour was applied  in \cite{10098904} and \cite{huo2022integrating} to map new UC problems to similar historical cases and retrieve relevant parameters for problem simplification. Similarly, \cite{10098904} and \cite{9524441} proposed to simplify security-constrained UC problems by applying FFN and CNN to predict the binary solutions. While these assisting approaches generally improve feasibility, achieving both solution feasibility and optimality simultaneously still remains a challenge. Furthermore, one major limitation of DL approaches lies in their limited ability to generalise to problem structure changes, such as the varying EV number in the EV JRS problem. Such changes can impact the number of decision variables, which can render DL approaches impractical since even minor variation in the problem structure would necessitate re-training. This leads to generation of separate labelled datasets for every problem configuration, thereby incurring significant labelling time from solving large number of MIP problem instances.

In this paper, we propose a DL-assisted strategy to accelerate the solution process of a day-ahead EV JRS stochastic optimisation problem with varying number of EVs. The approach involves training a CNN to predict the problem's binary variables, thereby reducing the solution search space and allowing a MIP solver to determine the remaining variables more efficiently. A padding mechanism is also designed to enable the CNN to handle changes in input and output sizes caused by varying number of EVs without the need for re-training. Furthermore, our JRS formulation also considers traffic conditions, a factor often neglected due to time-resolution discrepancies between the routing and scheduling side. Our contributions are listed as follows:

\begin{enumerate}
\item \textbf{Joint Routing and Scheduling Model with Dynamic Arc Travel Time:} We first formulate a MIP-based day-ahead EV JRS stochastic optimisation problem that accounts for traffic congestion. This formulation includes a modified Time-Space Network (TSN) with the addition of VCNs to represent more realistic EV travel times.  
\item \textbf{Adaptive DL to Solve Routing and Scheduling with Varying Number of EVs:} We then put forward a CNN to assist a MIP solver in solving the formulated problem. The CNN is trained to predict the binary variables, thus reducing the solution search space and enabling the remaining variables to be solved more efficiently. Additionally, a padding mechanism is included to adapt to changes in input and output sizes resulted from varying number of EVs. This eliminates the need for re-training, which is non-trivial due to difficulty in acquiring labelled data for a combinatorial challenging problem. 
\item \textbf{Validation on Grid and Transportation Network with Varying Number of EVs:}
To evaluate the generalisation capability of our approach, a case study was done on the IEEE 33-bus system combined with the Nguyen-Dupuis TN \cite{nguyen1984efficient} with number of EVs within the range of 20 to 100. Our approach achieved 97.8\% runtime reduction compared to unassisted Gurobi \cite{gurobi}, while retaining 99.5\% feasibility and deviating less than 0.01\% from optimality without the need of re-training.

\end{enumerate}

\noindent The remainder of the paper is as follows. Section \ref{Problem Formulation} presents the formulation of the day-ahead EV JRS stochastic optimisation problem. Section \ref{Methodology} details the proposed CNN-based acceleration strategy. Case studies and results are presented in Section \ref{result}, and the paper is concluded in Section \ref{conclude}.

\section{Problem Formulation} \label{Problem Formulation}

\begin{figure*}[!t]
\centerline{\includegraphics[width=\textwidth]{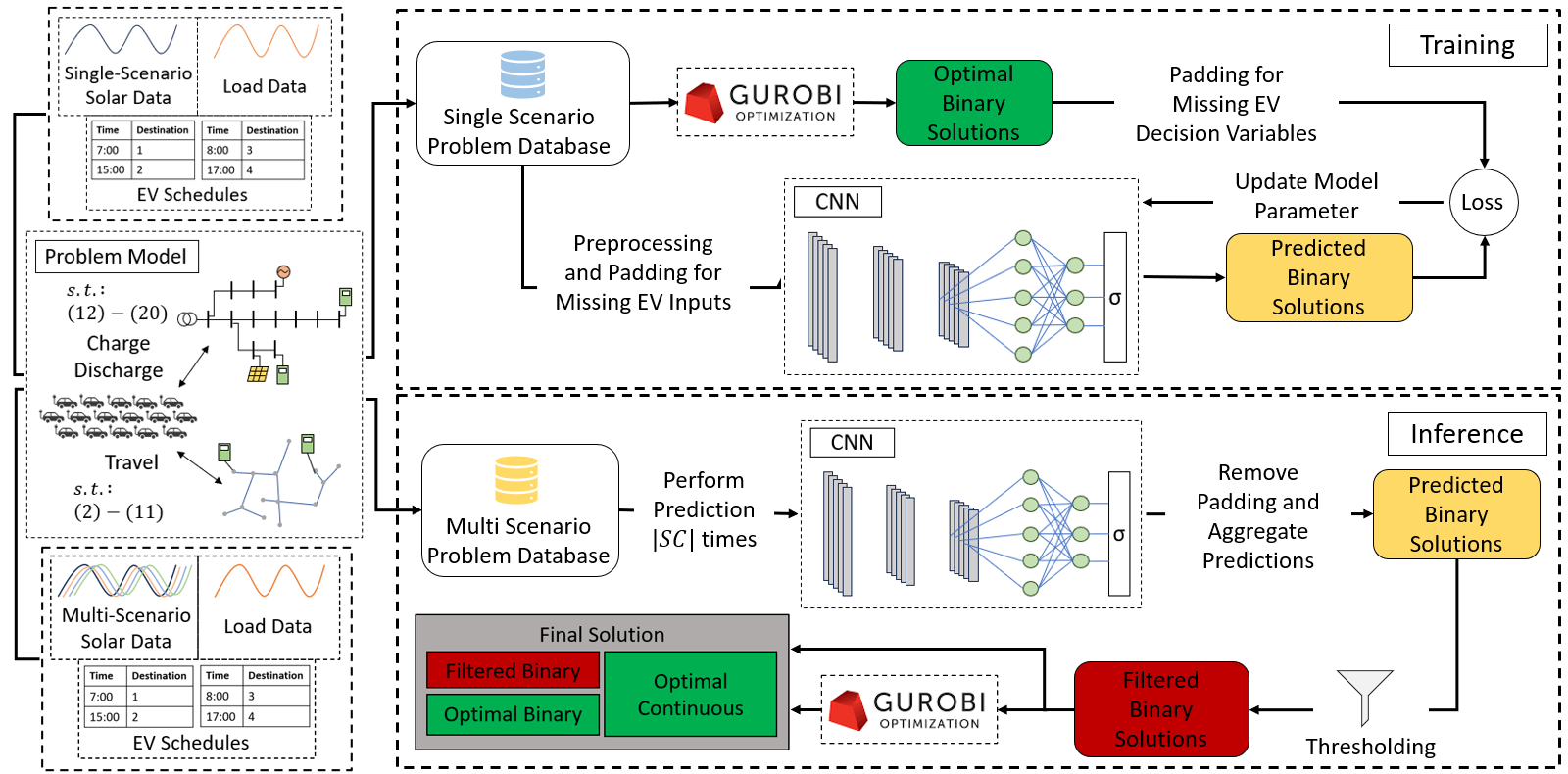}}
\caption{(Top) Overview of the proposed training and (Bottom) inference framework. The simplified problem model shows EVs' interaction with DN and TN, where \eqref{eq2}-\eqref{eq20} denote the constraints in Section \ref{Problem Formulation}.}
\label{FFN}
\end{figure*}

In this section, the formulation of a MIP-based day-ahead JRS stochastic optimisation problem for an EV fleet operating in a TN and a DN with solar PVs will be presented. When given a set of solar generation scenarios ($P^{pv_{max}}_{sc,p,t}$), load demand ($P^{L}_{b,t}$ \& $Q^{L}_{b,t}$) and EV job schedules ($\Xi$), the routing part of the problem aims to determine the optimal routes for each EV to complete its job schedules. Simultaneously, the scheduling part aims to schedule the distributed generator (DG) operations with the available solar generation to meet the load demand. Moreover, the EVs may also be routed to charge or discharge at appropriate time and location to provide grid support since EV charging stations introduce a coupling relationship between the DN and TN. A simplified problem model is provided in Fig.~\ref{FFN} to illustrate the EVs' interaction with the DN and TN.

\subsection{Modified Time-Space Network} \label{Modi TSN}

We adopt the TSN formulation from \cite{7024941} for the routing part of the JRS problem. In a TSN formulation, a TN is represented as a set of destination nodes interconnected by directed arcs. Each arc is modelled as a binary variable and every EV must select exactly one arc as its routing decision during each timespan. This will determine if the EVs will be travelling or idling (arcs with same source and target nodes). For trips that require more than one timespan, virtual nodes are added along the path to extend the trip times. However, it may not be appropriate to directly adopt the TSN in \cite{7024941}, as the study focuses on coordinating railway ESSs and assumes fixed travel times throughout the scheduling period. This assumption rarely holds for EVs operating within a TN due to varying traffic conditions. To address this, we modified the TSN formulation by introducing a VCN between each pre-existing destination and virtual nodes to create an alternate congestion path. By enabling and disabling the original arcs and the arcs leading to the VCNs, the travel time between any two pre-existing nodes may be extended or shortened based on the traffic conditions. However, it is important to always enable the arcs exiting the VCNs to allow the EVs to exit from the congestion state.

The problem objective and constraints are presented in \eqref{eq1}-\eqref{eq20}. The problem aims to minimise the DGs’ ($u\in G$) generation cost and EVs’ ($k\in K$) travelling and charging cost. This is reflected in the objective function \eqref{eq1}, where $sc \in SC$ is the set of problem scenarios, $t \in T$ is the set of timesteps, $s \in S$ is the set of timespans between each timestep, $NSA$ is the set of directed arcs where EVs would be non-stationary and travelling, and $CS$ is the set of charging stations,

\textit{Objective:}
    \begin{align}
        \begin{split} 
        \min\; 
        \sum_{sc\in SC} \sum_{u\in G} \sum_{t\in T} c^f (P^{g}_{sc,u,t})p_{sc} +
        \\ \sum_{sc\in SC} \sum_{k\in K} \sum_{ij\in NSA} \sum_{s\in S} c^t(I_{sc,k,ij,s})p_{sc} +
        \\ \sum_{sc\in SC} \sum_{k\in K} \sum_{i\in CS} \sum_{t\in T} [c^c(P^{c}_{sc,k,i,t}) - c^d(P^{d}_{sc,k,i,t})]p_{sc}. \label{eq1}
        \end{split}
    \end{align}
\noindent The first term in \eqref{eq1} indicates the total generation cost, where $P^{g}_{sc,u,t}$ is the generation power and $c^f$ is the generation cost coefficient. The second term indicates the total EV travelling cost with $c^t$ being the unit cost of travelling on each arc and $I_{sc,k,ij,s}$ being the the binary variable that indicates the chosen directed arc. Lastly, the third term indicates the net EV charging cost. It is determined by multiplying the cost and earning rates ($c^c$ and $c^d$) with the corresponding EV charging and discharging power ($P^{c}_{sc,k,i,t}$ and $P^{d}_{sc,k,i,t}$). Lastly,  since we account
for solar generation uncertainty by applying
scenario-based modelling, each term in \eqref{eq1} will be weighted by the occurrence probability of each scenario $p_{sc}$.

\subsection{Electric Vehicle Routing Constraints} \label{TC}
The problem constraints can be categorised into EV routing, EV power exchanging and grid operational constraints.  
\begin{gather}
        \begin{dcases*} \label{eq2}
            \sum_{ij\in CA} I_{sc,k,ij,s} = 1, & \text{if } $j_s = 1$ \\
            \sum_{ij\in NCA} I_{sc,k,ij,s} = 1, & \text{otherwise}
        \end{dcases*}  ,\forall{sc}, \forall{k}, \forall{s},
\\
        \sum_{ij\in CA\,\cap\,NCA} I_{sc,k,ij,s} = 1 ,\forall{sc}, \forall{k}, \forall{s}, \label{eq3}
    \\
    \begin{split}
    \sum_{ij\in A^{from}} I_{sc,k,ij,s+1} = \sum_{ij\in A^{to}} I_{sc,k,ij,s} , \\ \forall{sc}, \forall{k}, \forall [A^{from}, A^{to}], \forall{s=1,2,...|S|-1}, \label{eq4}
    \end{split}
    \\
    \sum_{ij \in A^{i+}_{\xi}} I_{sc,k_{\xi},ij,s_{\xi}} = 1, \forall{sc}, \forall{\xi \in \Xi = (k_{\xi}, i, s_{\xi})}, \label{eq5}
\end{gather}
\noindent  Constraints \eqref{eq2}-\eqref{eq5} specify the EV routing constraints. The modified TSN mentioned in Section \ref{Modi TSN} is modelled as~\eqref{eq2} and it indicates the arcs available for the EVs to travel on in the TN depending on the traffic condition. $CA$ and $NCA$ are the sets of directed arcs available during traffic congestion and non-congestion respectively, and $j_s$ is the binary that indicates the traffic condition. Next, \eqref{eq3} ensures that each EV can only travel on one arc per timespan. Moreover, \eqref{eq4} ensures the EV flow conservatism, where each EV should stop and start from the same node on timespan $s$ and $s+1$. $[A^{from}, A^{to}]$ is the set of pairs of directed arc sets that share the same starting and ending node respectively. Constraint \eqref{eq5} ensures each EV's origin node on each timespan is the scheduled destination if a job schedule exists, where each individual schedule $\xi \in \Xi$ specifies EV $k_{\xi}$ should be at node $i$ at timespan $s_{\xi}$, and $A^{i+}_{\xi}$ is the set of arcs that has node $i$ as starting point. 

\subsection{Electric Vehicle Power Exchanging Constraints} 
On the other hand, the EV power exchanging constraints are specified in \eqref{eq6}-\eqref{eq11}, 
\begin{gather}
    \begin{split}
    I^{c}_{sc,k,i,s+1} + I^{d}_{sc,k,i,s+1} \leq I_{sc,k,ii,s}, \\ \forall{sc}, \forall{k}, \forall{i\in CS}, \forall{s}, \label{eq6}
    \end{split}
    \\
    P^{m}_{sc,k,s+1} = \sum_{ij\in NSA} P^{move}(I_{sc,k,ij,s}), \forall{sc},\forall{k}, \forall{s},  \label{eq7}
    \\
    \begin{split}
        0 \leq P^{c}_{sc, k, i, t} \leq P^{ev,max}_{k}(I^{c}_{sc,k,i,t}),   \forall{sc}, \forall{k}, \forall{i\in CS}, \forall{t},
    \end{split} \label{eq8} 
    \\
    \begin{split}
        0 \leq P^{d}_{sc, k, i, t} \leq P^{ev,max}_{k}(I^{d}_{sc,k,i,t}),  \forall{sc}, \forall{k}, \forall{i\in CS}, \forall{t},
    \end{split} \label{eq9} 
    \\  
        E^{min}_{k} \leq E_{sc,k,t} \leq E^{max}_{k}, \forall{sc}, \forall{k}, \forall{t}, \label{eq10}
    \\
    \begin{split}
        E_{sc,k,t} = E_{sc,k,t-1} + \frac{24}{|T|}[(1-\eta)\sum_{i\in CS}P^{c}_{sc, k, i, t} - \\ (1+\eta)(\sum_{i\in CS}P^{d}_{sc, k, i, t} + P^{m}_{sc,k,t})] ,\forall{sc}, \forall{k}, \forall{t=2,...,t},
    \end{split} \label{eq11} 
    \end{gather} 
\noindent where \eqref{eq6} ensures that an EV can only charge or discharge if it chose to stay at a charging station. $I^{c}_{sc,k,i,s+1}$ and $I^{d}_{sc,k,i,s+1}$ are the binary variables that indicate whether an EV is charging or discharging respectively, and index $ii$ indicates the corresponding arc where EVs are stationary and will stay at charging station $i$. Constraint \eqref{eq7} indicates that an EV will only consume charge if it travels, where $P^{m}_{sc,k,s+1}$ is the power consumed by EVs when travelling and $P^{move}$ is the amount of energy consumed per arc travelled. Moreover, constraints \eqref{eq8}-\eqref{eq9} limit the maximum and minimum charging and discharging rates of each EV respectively, where $P^{c}_{sc, k, i, t}$ and $P^{d}_{sc, k, i, t}$ are the EVs' charging and discharging power. Similarly, \eqref{eq10} limits the maximum $E^{max}_{k}$ and minimum energy content $E^{min}_{k}$ of each EV. Last but not least, \eqref{eq11} ensures the energy balance of each EV, where $\eta$ is the EVs' charging and discharging loss.  The term $\frac{24}{|T|}$ is included to convert power to energy.

\subsection{Grid Operational Constraints}

The grid operational constraints \eqref{eq12}-\eqref{eq20} mainly includes the generation and DN constraints. Firstly, the generation constraints are specified by constraints \eqref{eq12}-\eqref{eq14}.
\begin{gather}
    P^{g,min}_{u}\leq P^{g}_{sc,u,t} \leq P^{g,max}_{u}, \forall{sc}, \forall{u}, \forall{t}, \label{eq12}
    \\
    Q^{g,min}_{u}\leq Q^{g}_{sc,u,t} \leq Q^{g,max}_{u}, \forall{sc}, \forall{u}, \forall{t}, \label{eq13}
    \\ 
    0 \leq P^{v}_{sc,p,t} \leq P^{pv_{max}}_{sc,p,t}, \forall{sc}, \forall{p} \in PV, \forall{t}. \label{eq14}
\end{gather}
\noindent Constraints \eqref{eq12}-\eqref{eq13} enforce the maximum and minimum active and reactive generation power of the DGs, where $P^{g}_{sc,u,t}$ is the actual active power generated and $Q^{g}_{sc,u,t}$ is the reactive power generated. Similarly, \eqref{eq14} enforces the maximum and minimum active power generation of each solar PV system, where $PV$ is the set of solar PVs and $P^{pv_{max}}_{sc,p,t}$ is the forecasted solar generation scenarios. The constraint also allows solar generation to be curtailed when necessary, where $P^{v}_{sc,p,t}$ represents the curtailed solar power.
    \begin{gather}
    -P^{f,max}_{l} \leq P^{f}_{sc,l,t} \leq P^{f,max}_{l}, \forall{sc},\forall{l} \in L, \forall{t}, \label{eq15}
    \\
    -Q^{f,max}_{l} \leq Q^{f}_{sc,l,t} \leq Q^{f,max}_{l}, \forall{sc},\forall{l} \in L, \forall{t}, \label{eq16}
    \\
    \begin{split}
        \sum_{u\in G_b} P^{g}_{sc,u,t} + \sum_{p\in PV_b} P^{v}_{sc,p,t} + P^{f}_{sc,l^{up}_{b},t} = \\ \sum_{l\in L^{down}_{b}} P^{f}_{sc,l,t} + P^{L}_{b,t} + \sum_{k} \sum_{i\in CS_b} (P^{c}_{sc, k, i, t} 
        \\ - P^{d}_{sc, k, i, t}) ,\forall{sc},\forall{b \in B},\forall{t}, 
    \end{split} 
    \label{eq17}
\\
    \begin{split}
        \sum_{u\in G_b} Q^{g}_{sc,u,t} + Q^{f}_{sc,l^{up}_{b},t} = \sum_{l\in L^{down}_{b}} Q^{f}_{sc,l,t}  
        \\+ Q^{L}_{b,t}, \forall{sc},\forall{b \in B},\forall{t},
    \end{split} 
    \label{eq18}
    \\
    \begin{split}
    V_{sc,b^{up}_{l},t} - (P^{f}_{sc,l,t}r_{l} + Q^{f}_{sc,l,t}x_{l})/V^{ref} =
    \\ V_{sc,b^{down}_{l},t},\forall{sc}, \forall{l} \in L, \forall{t},
    \end{split}
    \label{eq19}
    \\
    0.95 \ V^{ref} \leq V_{sc,b,t} \leq 1.05 \ V^{ref}, \forall{sc}, \forall{b}, \forall{t}. 
    \label{eq20}
    \end{gather}

\noindent On the other hand, \eqref{eq15}-\eqref{eq20} specify the line and grid power constraints. The maximum and minimum active and reactive line power are enforced by \eqref{eq15}-\eqref{eq16} with $L$ being the set of lines. $P^{f}_{sc,l,t}$ and $Q^{f}_{sc,l,t}$ are the actual active and reactive line power. The line flows are considered in this work using the LinDistFlow \cite{du2020interval} formulation and is represented with constraints \eqref{eq17}-\eqref{eq18}, also known as the power balance constraints. The power balance constraints essentially balance the generation power with the load power of each bus, where $G_b$, $PV_b$ and $CS_b$ are the sets of DGs, PVs and charging stations at bus $b$ respectively. $B$ is the set of DN buses, $L^{down}_{b}$ is the set of downstream lines of each bus and $l^{up}_{b}$ is the upstream line of each bus. Moreover, $P^{L}_{b,t}$ and $Q^{L}_{b,t}$ are the active and reactive bus load. Constraint \eqref{eq19} models the voltage balance, where $r_l$ and $x_l$ are the line resistance and impedance respectively, $b^{up}_{l}$ and $b^{down}_{l}$ are the up and downstream buses that line $l$ is connected to respectively, and $V^{ref}$ represents the slack bus voltage. Lastly, \eqref{eq20} ensures the bus voltage is within 5\% of $V^{ref}$, where $V_{sc,b,t}$ is the bus voltage. 

\section{Methodology} \label{Methodology}

\begin{figure*}[!t] 
    \centering
    \includegraphics[width=0.95\textwidth]{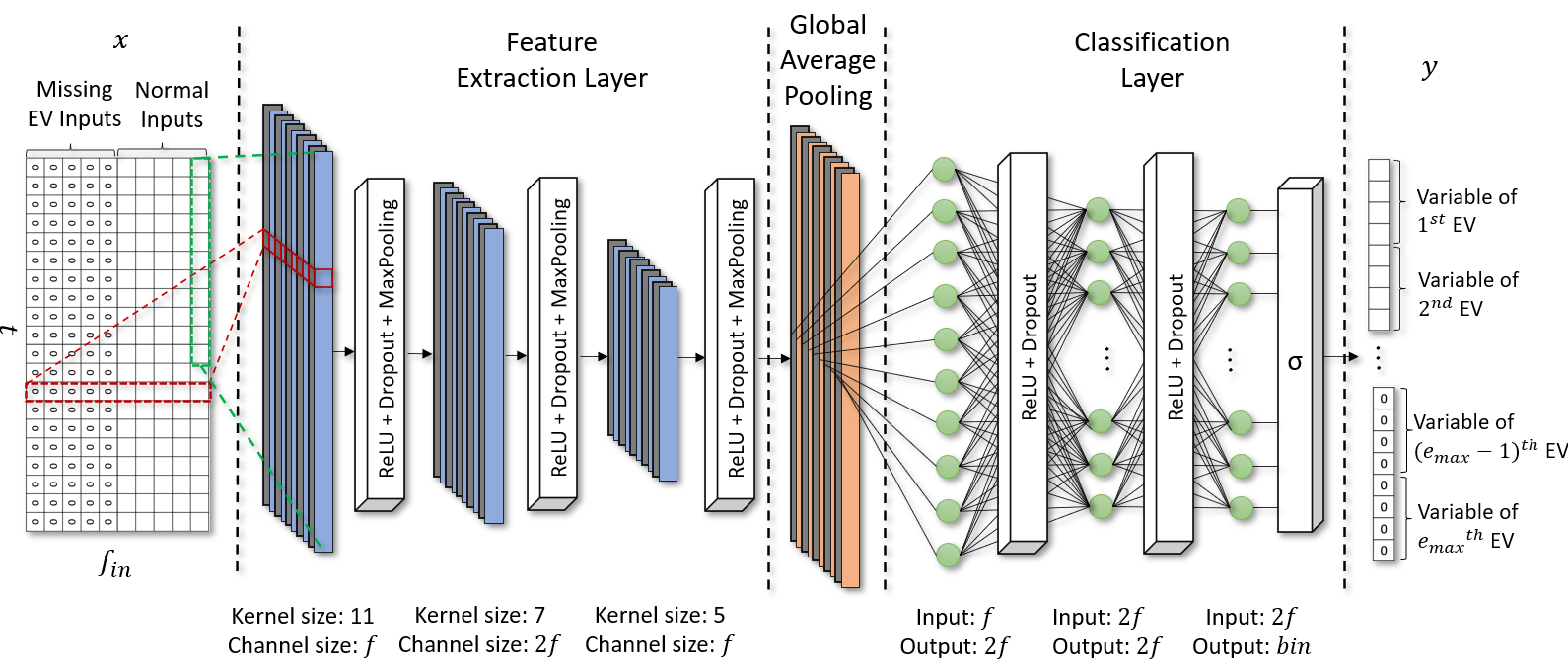}
    \caption{Structure of CNN in this work, where $e_{max}$ is the estimated maximum number of EVs and the input and output shapes shown are compressed.}
    \label{CNN}
\end{figure*}

\subsection{Dataset} \label{DataGen}

As outlined in Section \ref{introduction}, the goal is to assist Gurobi to solve the formulated problem by training a CNN to predict the binary decision variables. The proposed framework is summarised in Fig. \ref{FFN}. Supervised learning is employed as the training methodology and a problem database is first constructed to generate the labelled training dataset. A large number of different input data (solar generation, bus load and the EV job schedules) will first be required to build a diverse problem database. However, we synthetically generated these parameters using the following steps due to a lack of data. 

We first generate the solar generation scenarios by constructing a probability density function (PDF) using historical data. Realistic profiles are then sampled from it using Monte Carlo scenario generation technique. Next, since the bus load tends to exhibit lesser variations, we generate them by scaling a base load profile with random coefficients sampled from $[0.9,1.1] \in \mathbb{R}$. Finally, we randomly generate job schedules for each of the EV.

We then proceed to construct the training problem database with the generated parameters. Since scenario-based modelling is employed to consider the solar generation uncertainty, each problem instance will require $|SC|$ samples of solar generation scenarios, and one set of bus load and EV job schedules. Each problem instance can then be labelled by solving them with Gurobi to obtain the optimal binary solutions. However, the labelling time can become prohibitively long, especially if a high scenario number is used to better capture the solar generation uncertainty. Thus, we proposed to train the CNN with deterministic version of the problem instead. During inference, the CNN can instead be applied to predict the binary variables of each scenario independently and aggregating them to form the complete binary solution. This approach significantly reduces the labelling time by avoiding the need to solve large number of the full stochastic problem~\eqref{eq1}.

\subsection{Deep Learning Training}

Based on the constructed training dataset, the input for our CNN is the solar-load-schedule combination used to construct each problem instance, while the output is the corresponding optimal binary solution. Both the solar generation \(SG \in \mathbb{R}^{\,1\times p \times t}\) and active-reactive bus load \(PQ \in \mathbb{R}^{\,(2 \times b)\times t}\) are naturally time-series, allowing them to be directly taken as input, where \(p\) is the number of PV panels, \(b\) is the number of DN buses, and \(t\) is the number of timesteps. In order to incorporate the EV job schedules into the input, a time-series representation was created. Specifically,  an array of length \(t\) was generated for each EV, where the scheduled destination nodes were set to their corresponding scheduled timestep in the array. For instance, if an EV was scheduled to be at node $a$ at timestamp 8, the value at index 8 of the array would be set to $a$. Therefore, the EV job schedules can be represented as \(\Xi \in \mathbb{R}^{\,e \times t}\), where \(e\) is the number of EVs. Consequently, the training problem can be framed as a time-series multi-label classification problem. The inputs are then normalised before splitting the dataset into training (90\%) and validation (10\%) sets. A separate set of testing data with the full stochastic problem~\eqref{eq1} is then generated. More details regarding the training and testing dataset will be discussed in Section \ref{result}.

Fig. \ref{CNN} illustrates the CNN architecture used in this work. Although CNNs are primarily used for processing images, they can be equally effective when applied to multivariate time-series data. This is due to their ability to capture spatial structures, therefore capturing the dependencies across the time and feature dimension of the data\cite{wang2017time}. However, a challenge arises from the potential change in number of EVs in the problem, which would affect the input and output sizes, thus necessitating re-training. This renders the approache impractical, as constructing a separate dataset for every problem configuration would incur significant labelling time from solving large number of the formulated problem instances.

To address this, we proposed a padding mechanism as illustrated in Fig. \ref{CNN}, to eliminate the need for re-training. One observation that could be made in the formulated problem is the number of binary variables scales with the number of EVs. Therefore, the idea is to first estimate the maximum number of EVs expected in the problem and designing the CNN accordingly. For problem instances with fewer number of EVs, the inputs and outputs of each missing EV are then padded with zeros to maintain the input and output sizes. During inference, the inputs of each scenario are then similarly padded before making the predictions. The padded zeros are then removed from each of the prediction before aggregating all remaining predictions to form the full binary solution.

\subsection{Post-processing Layer} \label{Post-processing Layer}
\begin{algorithm}[!b]
\caption{Inference framework}\label{alg:cap} \textbf{Input:} $SG \in \mathbb{R}^{\,|SC|\times p \times t}$, $PQ \in \mathbb{R}^{\,(2 \times b)\times t}$ and $\Xi \in \mathbb{R}^{\,e\times t}$ \\
\textbf{Output:} Solution of problem \eqref{eq1}
\begin{algorithmic}[1]
\State $\hat{y} \gets [~]$
\For{each $SC$}
\State $\hat{y}_{sc} \gets CNN(SG_{sc},\,PQ,\,Padding(\Xi))$
\State remove padding from $\hat{y}_{sc}$
\State $\hat{y} \gets Concatenate(\hat{y},\,\hat{y}_{sc})$
\EndFor
\For{each variable in $\hat{y}$}
\If{$\bar{p_t}^{0} \geq \hat{y} \geq \bar{p_t}^{1}$}
\State $Dispose(\hat{y})$
\Else
\State $Round(\hat{y})$
\EndIf
\EndFor
\\
\Return $Gurobi(SG,\,PQ,\,\Xi,\,\hat{y})$
\end{algorithmic}
\end{algorithm}

Once the CNN is trained, it can then be used to assist Gurobi in solving the formulated problem. While Gurobi can enforce solution feasibility and optimality, a feasible prediction will still be required for Gurobi to determine the remaining variables. As highlighted in Section \ref{introduction}, DL approaches often cannot enforce problem constraints and attempts on repairing infeasible predictions often compromises solution quality. Thus, we proposed to apply a thresholding filter to filter out incorrectly predicted bits that could potentially cause infeasibility instead. The thresholds ($\bar{{p_t}}^{1/0}$) used are the mean prediction probabilities (PPs) of each class. The PP of each variable is calculated with \eqref{eq24} proposed in \cite{ridnik2021asymmetric}, where ${p_t}^{1/0}$ are the PPs of each binary variable, $\hat{y}$ is the prediction of the CNN and $y$ is the actual label. 
\begin{equation}\label{eq24}
    p_t^{1/0} = \hat{y} \cdot 1_{\{y=1\}} + (1-\hat{y}) \cdot 1_{\{y \neq 1\}}.
\end{equation}
This method involves extracting predictions with probabilities above or below a threshold, leveraging the fact that Sigmoid output essentially represents the probability of a prediction being 1. The calculated $\bar{{p_t}}^{1/0}$ should enable us to maximise both the number of variables extracted to assist Gurobi and the feasibility of the filtered binary solution. The filtered binary variables can then be used to reduce the solution search space and allow Gurobi to determine the remaining binary and continuous variables more efficiently. The psuedocode of the inference framework is provided in Algorithm \ref{alg:cap}. However, infeasibility can still occur occasionally. In those cases, we increase $\bar{{p_t}}^{1}$ by 0.1 and proceed to re-optimise the problem. 

\section{Case Study and Result} \label{result}

\begin{figure}[!t]
\centerline{\includegraphics[width=\linewidth]{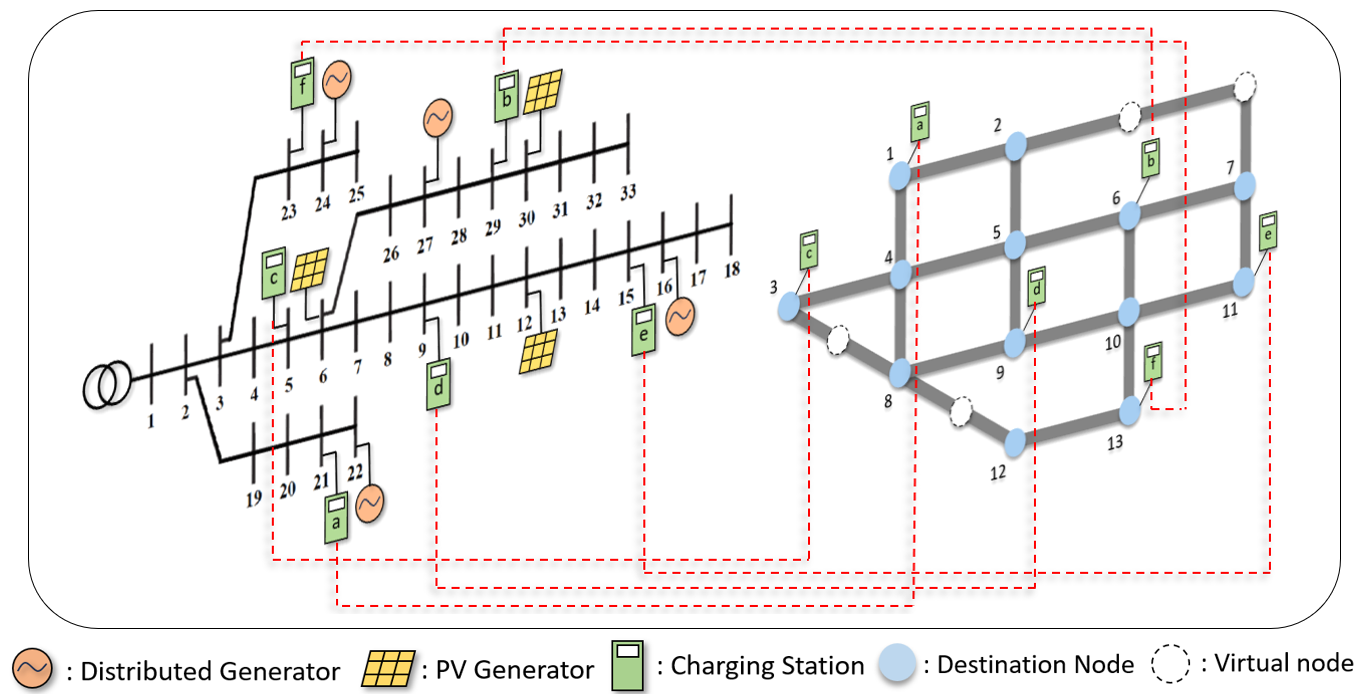}}
\caption{IEEE 33-bus system and Nguyen-Dupuis transportation network.}
\label{network}
\end{figure}

\subsection{Simulation Setup} \label{CS}
The formulated problem and the CNN model were implemented using Python Gurobi and PyTorch. All computations were performed on a system with Intel® Core™ i9-12900K Processor, 128 GB of RAM and NVIDIA RTX™ A5000 GPU. We validated the proposed approach by conducting a case study on the IEEE 33-bus system coupled with the Nguyen-Dupuis TN as shown in Fig.~\ref{network}. The input data used to construct the problem database outlined in Section \ref{Methodology} are listed as follows. Historical solar generation data, obtained from Solcast API Toolkit \cite{solcast}, was used to generate the solar generation PDF. The number of scenarios was set to 5, each with equal occurrence probability. The base load curve was derived by scaling the IEEE 33-bus system bus load data with an interpolated coefficient vector provided in \cite{TRIVINOCABRERA2019113}. Lastly, we assume that each EV's job schedule includes a random starting and destination node at the beginning and end of the day, along with two other shifts at two random nodes and times throughout the day. Therefore, each EV's job schedule specifies that an EV will first have a random start-destination node pair selected from $[(1,11), (1,13), (3,11), (3,13)]$. The EVs will then have a shift at a random time between 6:00 AM and 12:00 PM at a random node selected from $[6,7,9,10,12]$, followed by a second shift between 2:00 PM to 10:00 PM at a random node from $[2,4,5,8]$ if the starting node was $1$. Conversely, the first shift will be from $[2,4,5,8]$ and the second shift from $[6,7,9,10,12]$ if the starting node was $3$. We sampled a total of 1,000 solar profiles, 200 bus load profiles and 20 sets of job schedules for each EV for building the problem database mentioned in Section \ref{Methodology}. Code for reproducing experiments will be available at https://github.com/YapJunKang/EVJRS.

\subsection{Training and Validation}

Although the proposed framework is designed to accommodate changes in input and output sizes caused by varying number of EVs, it is not necessarily capable of providing feasible or high-quality solutions if the training dataset lacks sufficient problem variations. Thus, we aim to evaluate the generalisation capability of the CNN when exposed to datasets with varying degrees of unseen scenarios. We first set the number of EVs to the range of 20 to 100. We then constructed four different training datasets, with each containing 800 samples of problem instances with different multipliers of EV numbers as shown in Fig. \ref{dataset}. Four different CNNs are then trained on each of the dataset. Subsequently, a separate testing dataset was constructed with EV numbers that were not present in any of the training dataset to evaluate the generalisation performance of each CNN. We set a MIPGap of 0.1\% or 120 minutes timeout in Gurobi when labelling the datasets.
 
When evaluating the prediction performance of the trained CNNs, class-specific accuracies ($Acc_0$ and $Acc_1$) and mean average precision (mAP) are calculated instead of the overall accuracy due to the significant class imbalance issue. While these metrics provide insights into the CNNs' prediction performances, they do not fully capture its generalisation capability when solving the formulated problem. Thus, the trained CNN is further assessed using the following metrics to evaluate their performance in assisting Gurobi to solve the formulated problem. The first metric is the average percentage of computational time reduction ($\bar{r}$). The second metric is the average percentage of solution optimality loss ($\bar{l}$) and the third metric is the percentage of test samples where the proposed approach returned a feasible solution ($feas$). Lastly, in order to have a fairer comparison, we assume all infeasible samples will be solved by Gurobi when calculating $\bar{r}$.

\subsection{Results and Discussion}
\begin{figure}[!t]
\centerline{\includegraphics[width=\linewidth]{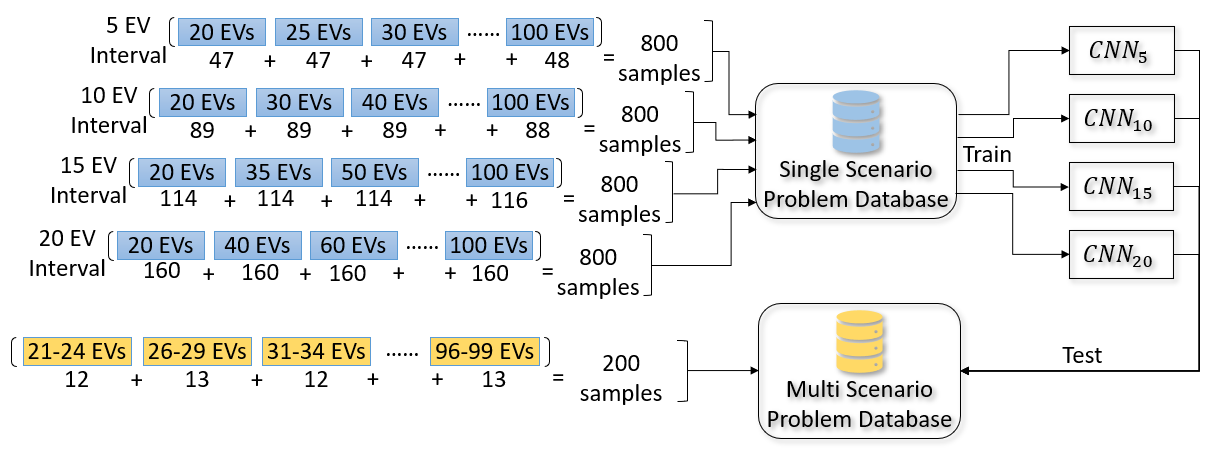}}
\caption{Number of EVs of problem instances in training and testing dataset.}
\label{dataset}
\end{figure}

A total of four CNNs have been trained on each of the single scenario interval dataset mentioned in Fig. \ref{dataset}. The hyper-parameters of each CNN are tuned with Optuna \cite{optuna_2019} to achieve the lowest training loss. All the final model hyper-parameters after tuning will be provided in the link mentioned previously. Once trained, all CNNs are tested on the same set of 200 samples of the full stochastic problem with number of EVs unseen to all CNNs. The CNNs' prediction performances are reported in Table \ref{table1}. All CNNs had shown similar performances, with $CNN_{15}$ performing slightly better than others, following by $CNN_{10}$, $CNN_{5}$ and $CNN_{20}$. This comes as a surprise as models trained on dataset with higher degree of unseen scenarios should exhibit worse generalisation capability since the models will have to make more approximations to fill in those unseen scenarios.  

\begin{table}[!b] 
\caption{Performance Metrics of CNNs Trained on the Different Interval Dataset.}
    \begin{center} \label{table1}
    \begin{tabular}{ccccc}
    \hline
    \textbf{Performance}&\multicolumn{4}{c}{\textbf{CNN Solvers}} \\
    \cline{2-5} 
    \textbf{Metrics (\%)} & \textbf{$CNN_5$}& \textbf{$CNN_{10}$}& \textbf{$CNN_{15}$}& \textbf{$CNN_{20}$}\\
    \hline
    \textbf{$Acc_0$} & 99.88& 99.87& 99.87& 99.85 \\
    \textbf{$Acc_1$} & 59.62& 60.43& 61.12& 60.44 \\
    \textbf{$mAP$} & 74.00& 75.00& 75.00& 73.00\\
    \textbf{$\bar{p}_t^0$} & 99.58& 99.57& 99.58& 99.56 \\
    \textbf{$\bar{p}_t^1$} & 71.64& 70.87& 70.93& 70.02 \\
    \hline
    \textbf{$\bar{r}~$} & 97.83 & 95.81& 91.48& 75.59\\
    \textbf{$\bar{l}~$} & -0.01& 0.06& 0.29& 0.52\\
    \textbf{$feas$} & 99.50& 97.50& 95.50& 77.00\\
    \hline
    \end{tabular}
    \end{center}
\end{table}

\begin{figure}[!t]
\centerline{\includegraphics[width=0.9\linewidth]{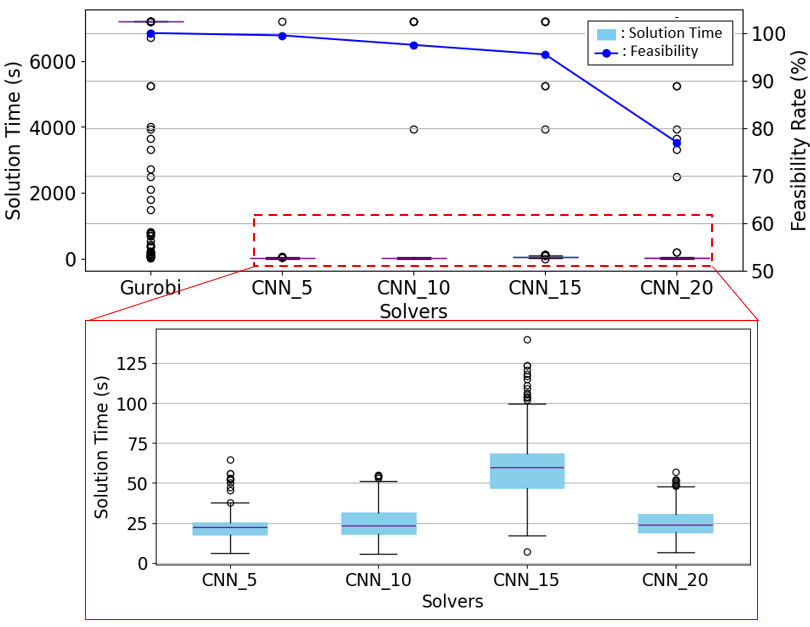}}
\caption{Solution time and feasibility rate of each CNN trained on different EV interval datasets.}
\label{performance}
\end{figure}

Although $CNN_{15}$ had shown the best prediction performance, this does not necessary translate to it being better at assisting Gurobi to solve the formulated problem. The performances of each CNN at assisting Gurobi to solve the test samples are summarised in Table \ref{table1}. Firstly, all CNNs are capable of assisting Gurobi to reduce the computational time by more than 91\% (except for $CNN_{20}$) on average, while losing less than 0.6\% solution optimality. This is non-trivial as this allows system operators to make quick decisions for time-sensitive tasks without the need of sacrificing solution quality. However, one limitation of the proposed approach is that there are no guarantees to feasibility as shown in Table~\ref{table1}, where we were only able to achieve 99.5\% feasibility rate with $CNN_5$. Moreover, as the EV interval in the dataset increases, we can see a noticeable drop in solution feasibility, and consequently a drop in computational time reduction and solution quality. This makes sense since there are more approximations that the CNNs will have to make to fill in the unseen scenarios. However, there are possible solutions to improve the feasibility rate of the proposed approach. The first is to further tune $\bar{{p_t}}^{1/0}$ to filter out more incorrectly predicted bits. The second is to further reduce the EV interval in dataset. However, since the problem is combinatorial challenging, it may be impractical to obtained more labelled data. Thus the former approach is more encouraged. To further show the significance of the proposed approach, we plotted the actual computational time of each test sample for each solver in seconds as shown in Fig.~\ref{performance}. The computational time of Gurobi without any assistance has large standard deviation and some samples can take unacceptably long to solve, while $CNN_5$ only required less than a fraction of the time for majority of the test instances. Note that although solution time median of $CNN_{15}$ is higher, it has better $\bar{r}$ than $CNN_{20}$ overall due to having higher $feas$.

\subsection{Ablation Study -- Padding Mechanism}

In this section, we will discuss the advantage of the proposed padding mechanism. As highlighted in Section \ref{Methodology}, incorporating the padding mechanism into the CNN architecture is recommended to enable generalisation to varying number of EVs in the problem and avoid re-training. This is because re-training will significantly increase the labelling time since new training dataset will be required for every new model. To demonstrate this, we carry out the following study to approximate the time required to generate the labelled training datasets for CNNs with and without the padding mechanism for problems with number of EVs within the range of 20 to 100 (i.e. 81 different numbers of EVs). A summary of the dataset details required for training the CNNs with and without padding is provided in Table \ref{datadetail}. For the CNN with padding, $CNN_5$ described in the previous section was chosen. In contrast, the CNN without padding requires a separate model to handle problems with each number of EVs. Assuming only 400 samples are needed to train each CNN without padding to achieve similar performance shown in Table \ref{table1}, this will result in a total of $81$\,datasets\,$\times\,400$\,samples\,$=32,400$ samples. To estimate the labelling time, we first solve 10 random problem instances for each EV number using Gurobi and calculated the mean solution time for each EV number. We then multiply the required number of samples for each CNN with the corresponding mean solution time to approximate the total labelling time. It can be observed that the time required to label all the dataset for CNN without padding is significantly higher, highlighting the importance of the padding mechanism when training a DL model to assist Gurobi in solving the formulated problem.

\begin{table}[t] 
\caption{Samples and Labelling Time for CNN with and without Padding mechanism.}
    \begin{center} \label{datadetail}
    \begin{tabular}{ccc}
    \hline
    \textbf{Dataset Attribute} & \textbf{Non-Padding}& \textbf{Padding} \\
    \hline
    \textbf{EV Number} &${e\:|\:e \in \{20,21,...,100\}}$ & ${5e\:|\:e \in \{4,5,...,20\}}$ \\
    
    \textbf{Total Samples} & $81 \times 400$& $1 \times 800$ \\

    \textbf{Labelling Time (h)} & 9948.88& 234.57 \\
    \hline
    \end{tabular}
    \end{center}
\end{table}

\subsection{Limitation and Future Work}
The results show that the proposed DL approach is capable of assisting Gurobi in solving the EV JRS problem efficiently. This makes the problem more tractable when applied to time-sensitive tasks and enables system operators to obtain high quality solutions in a timely manner. However, a key limitation of the approach is that it does not eliminate the computational burden, but instead redistributes it to the labelling process during training. This limitation can potentially hinder the scalability of the approach.

This work attempts to address this limitation by introducing a padding mechanism that enables the CNN to generalise to problems with varying number of EVs. By improving the generalisation capability of the DL model, the number of labelled problem instances needed for training can be reduced, as less frequent re-training will be necessary, thereby reducing the overall computational burden. Therefore, a promising direction for future work can be to further improve the DL model's generalisation capability in other aspects of the problem, such as the potential changes in DN and TN topology. 

Another solution that can potentially further alleviate the overall computational burden is the use of reinforcement learning (RL). RL algorithms are known to be capable of learning without needing any prior knowledge provided by expert agents, and can sometimes even out-perform them. This essentially eliminates the need for dataset labelling, thereby significantly reducing the computational burden. Consequently, RL presents itself as a promising future work direction. However, careful design of algorithms and reward functions should be carried out as RL often suffers from challenges such as poor sample efficiency and sparse reward function, which can make the training process difficult and inefficient \cite{QIU2023113052}.

\section{Conclusion} \label{conclude}

In this work, we introduced a DL-assisted framework to solve a MIP-based day-ahead JRS stochastic optimisation problem for an EV fleet operating in a TN and DN equipped with RESs. We included a modified TSN  in the formulation to account for traffic conditions. A CNN was trained to assist MIP solvers by predicting the binary variables, thereby reducing the solution search space and allowing the remaining variables to be solved more efficiently. To address the challenge posed by possible varying number of EVs in the problem, a padding technique was applied to handle changes in input-output sizes. In a case study on the IEEE 33-bus system and Nguyen-Dupuis TN with 20 to 100 EVs, we reduced runtime by 97.8\% when compared to Gurobi while retaining 99.5\% feasibility and deviating less than 0.01\% from the optimal solution without requiring re-training. 

\bibliography{references}{}
\bibliographystyle{IEEEtran}

\end{document}